\documentclass[11pt,twoside]{article}

\usepackage{asp2014}

\aspSuppressVolSlug
\resetcounters

\begin{document}

\title{PyMUSE: a Python package for VLT/MUSE data}

\author{Ismael~Pessa,$^1$ Nicolas~Tejos,$^2$ and Cristobal~Moya$^1$
\affil{$^1$Instituto de Astrof\'isica, Pontificia Universidad Cat\'olica de Chile, Vicu\~na Mackenna 4860, Santiago, Chile\\
}
\affil{$^2$Instituto de F\'isica, Pontificia Universidad Cat\'olica de Valpara\'iso, Casilla 4059, Valpara\'iso, Chile}
}

\paperauthor{Sample~Author1}{Author1Email@email.edu}{ORCID_Or_Blank}{Author1 Institution}{Author1 Department}{City}{State/Province}{Postal Code}{Country}
\paperauthor{Sample~Author2}{Author2Email@email.edu}{ORCID_Or_Blank}{Author2 Institution}{Author2 Department}{City}{State/Province}{Postal Code}{Country}
\paperauthor{Sample~Author3}{Author3Email@email.edu}{ORCID_Or_Blank}{Author3 Institution}{Author3 Department}{City}{State/Province}{Postal Code}{Country}

\begin{abstract}
This is a companion Focus Demonstration article to the PyMUSE python package, demonstrating its usage and utilities for VLT/MUSE data analysis, that include a wide range of options for spectra extractions, the creation of different types of images, compatibilities with some commonly used software for astronomical data analysis, among others. PyMUSE is an open-source software and can be found on Github for free use and distribution.
\end{abstract}

\section{Introduction}
Integral field spectroscopy has demonstrated to be a powerful tool to study the physical properties of extended astronomical objects (e.g. kinematics and chemical distribution) or to search for extremely faint objects (e.g. with low continuum level but with emission lines), among others.  MUSE \citep{muse2014} is the first large integral field spectrograph ever installed at an 8-meter telescope. It uses 24 spectrographs to separate light into its component colors to create both images and spectra of its field of view, which has a size of $1$'$\times1$' and is sampled at 0.2"/pixel. 
We present PyMUSE, a python package mostly aimed at analysis of VLT/MUSE datacubes. The package is optimized to extract 1-D spectra of arbitrary spatial regions within the cube and also for producing images using photometric filters and customized masks. It is intended to provide the user the tools required for a complete analysis of a MUSE data set. For completeness, we mention that other Python libraries for similar purposes exist, including MPDAF \citep{MPDAF} and LSDCat \citep{LSDCat}.

\section{PyMUSE subroutines}
\subsection{Spectrum extraction}
\begin{figure*}[!h]
\centering
\includegraphics[scale=0.4]{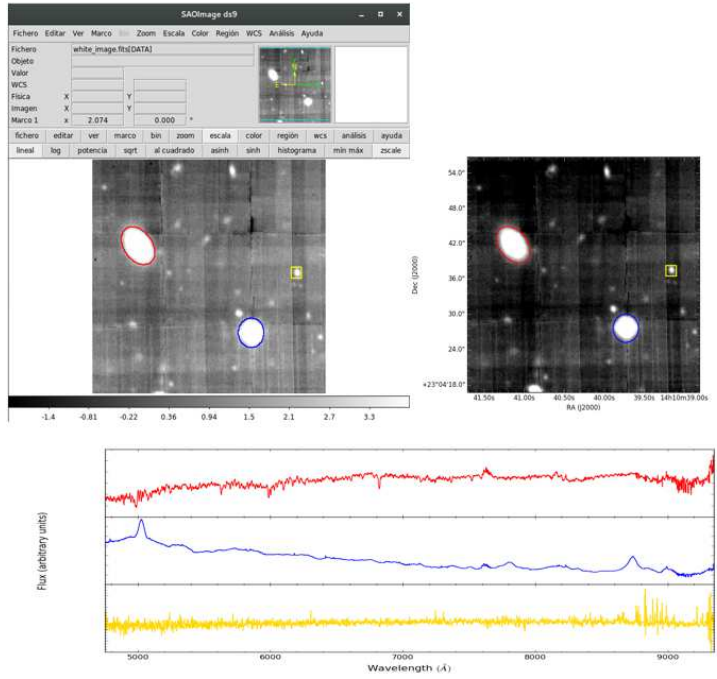}
\caption{Example of defining apertures using a DS9 region file. \textit{Upper-Left}: Original regions defined in DS9. \textit{Upper-rigth}: PyMUSE interface when the region file is given as input. \textit{Lower}: Output spectra of the three regions extracted using the weighted sum mode based on the white profile. }
\label{regions}
\end{figure*}

The main goal of PyMUSE is versatility at getting a spectrum. The aperture of extraction for a given source can be defined by a variety of manners:
\begin{itemize}
\item Elliptical set of parameters ($x_{c}$, $y_{c}$, $a$, $b$, $\theta$).
\item DS9 regions.
\item Interactive canvas.
\end{itemize}
Figure \ref{regions} shows an example of how DS9 can be used to define apertures to extract a spectrum.
Once the aperture is defined, the combination of the spaxels\footnote{We refer to spaxels as the pixels defined in the spatial direction.}  inside the aperture can be done in a set of different ways:
\begin{itemize}
\item Sum, median and mean.
\item Sum of the brightest X\% of the spaxels inside the aperture.
\item Weighted sum of the spaxels by a bright profile, either obtained from the MUSE white\footnote{The white image contains the sum of all the wavelengths of the datacube.} image or from a Gaussian profile. 
\item Weighted sum by the inverse of the variances (spatially or spectral).
\item Combinations of the last 2 points.
\end{itemize}

\subsection{Image creation}
\label{im}
Users can use PyMUSE routines to create different kinds of images according to their needs.
\begin{itemize}
\item  Filtered images to do photometry, by convolving the spectral axis of the datacube with SDSS or Johnson photometric filters is possible (customized filters can also be used). This feature can be very useful to characterize in detail sources in the field.
\item "Emission line filter" at different redshifts can be useful to detect faint star-forming galaxies that could be not detectable in the white image and can help the user to perform a complete analysis.
\item Smoothed and Masked images are also supported. The mask can be defined simply by a DS9 region file.
\end{itemize}

\subsection{Compatibilities with external software}
\subsubsection{DS9}
A useful PyMUSE utility for a  systematic analysis is to receive as an input a DS9 region file containing the apertures of interest. The user can select a set of regions and immediately save the corresponding 1-D spectra for a posterior analysis.
\subsubsection{SExtractor}
The user can perform a systematic search of sources using SExtractor \citep{sextractor}. In this case, the generated catalog can be used as a PyMUSE input to extract the 1-D spectra. The user can use this feature even if SExtractor is run in an image different than the MUSE image (astrometry should be checked in this case). Using SExtractor in a filtered image described in Section \ref{im} permit the user to simultaneously build a photometric and spectroscopic catalog of sources in the field.
\subsection{Redmonster}
Redmonster software \citep{hutchinson2016} is a set of Python utilities for redshift measurement and classification of 1-D spectra by performing $\chi^{2}$ minimization respect to a set of modulated theoretical models. The user can set the output spectra of PyMUSE to be directly used as Redmonster inputs. Figure \ref{rm} shows the final redshifts and classification obtained by Redmonster for the spectra used in the example above. 
\articlefigure{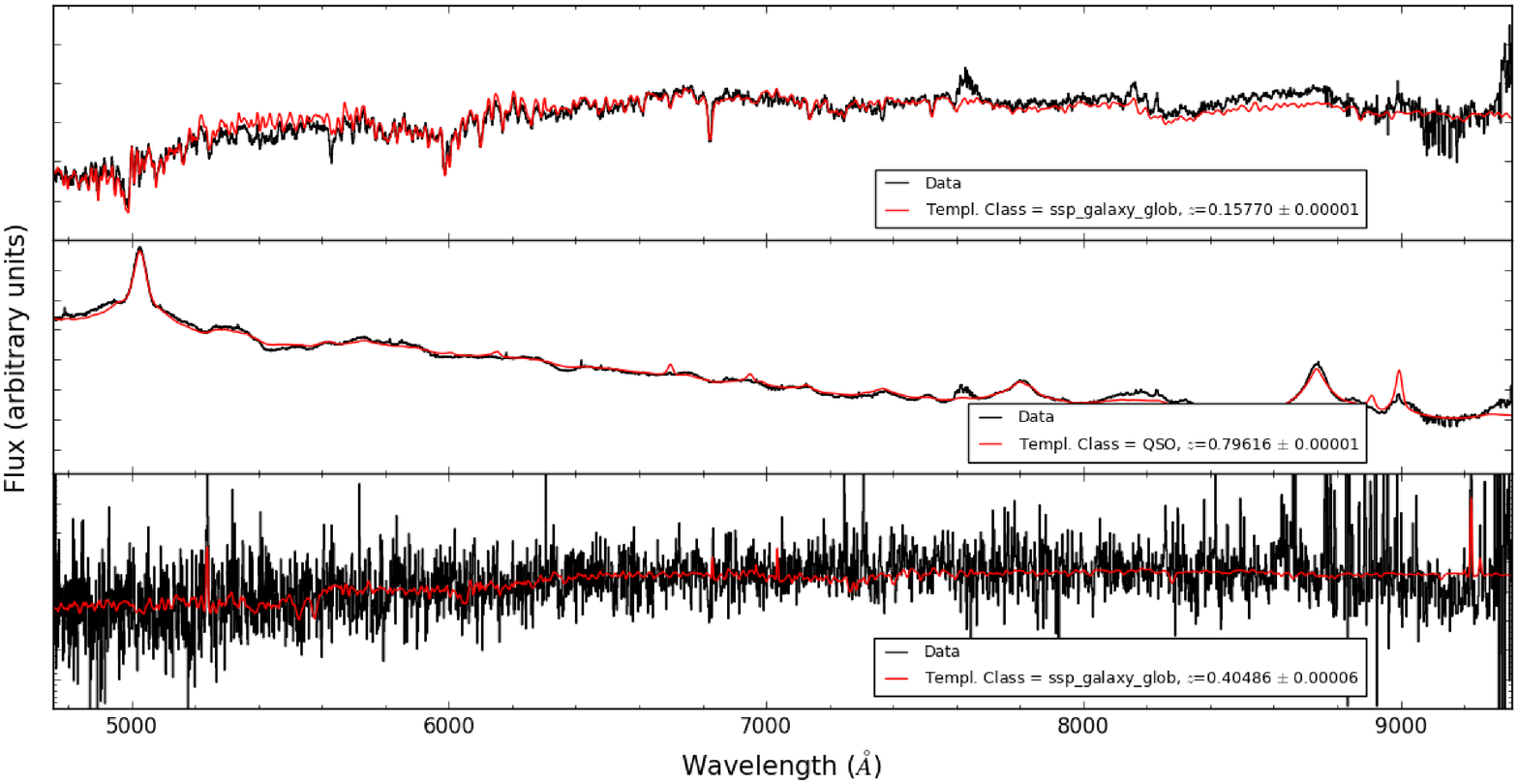}{rm}{Redmonster outputs for the 1-D spectra obtained from the apertures shown in Figure \ref{regions}.}

\section{Summary}
PyMUSE has shown to be a reliable tool for integral field spectroscopy data analysis, particularly for VLT/MUSE datacubes. It is well suited for deep field sources catalog and provides a set of routines focused for a systematic analysis to their data. It is intended to be used together with some other commonly used software for data analysis but it can still be useful to explore a data set by itself. Integral field spectroscopy is being more requested every year and the development of analysis tools will help the scientific community to make a more efficient use of their data. Currently, PyMUSE is being developed by a small team of scientist and we would appreciate contributors that help us on the improvement of this tool. Its current version can be found and downloaded from \url{https://github.com/ismaelpessa/PyMUSE}.

\acknowledgements We acknowledge support from {\it CONICYT PAI/82140055.} We also would like to thank the PYTHON programming language and the open-source community, in particular, the Astropy collaboration \citep{astropy} and also thank contributors of SciPy, Matplotlib, and Numpy.

\bibliography{F5.bib}

\begin{thebibliography}{}
\expandafter\ifx\csname natexlab\endcsname\relax\def\natexlab#1{#1}\fi
\expandafter\ifx\csname url\endcsname\relax
  \def\url#1{\texttt{#1}}\fi
\expandafter\ifx\csname urlprefix\endcsname\relax\def\urlprefix{URL }\fi
\providecommand{\eprint}[2][]{\url{#2}}

\bibitem[{{Astropy Collaboration} et~al.(2013){Astropy Collaboration},
  {Robitaille}, {Tollerud}, {Greenfield}, {Droettboom}, {Bray}, {Aldcroft},
  {Davis}, {Ginsburg}, {Price-Whelan}, {Kerzendorf}, {Conley}, {Crighton},
  {Barbary}, {Muna}, {Ferguson}, {Grollier}, {Parikh}, {Nair}, {Unther},
  {Deil}, {Woillez}, {Conseil}, {Kramer}, {Turner}, {Singer}, {Fox}, {Weaver},
  {Zabalza}, {Edwards}, {Azalee Bostroem}, {Burke}, {Casey}, {Crawford},
  {Dencheva}, {Ely}, {Jenness}, {Labrie}, {Lian Lim}, {Pierfederici},
  {Pontzen}, {Ptak}, {Refsdal}, {Servillat}, \& {Streicher}}]{astropy}
{Astropy Collaboration}, {Robitaille}, T.~P., {Tollerud}, E.~J., {Greenfield},
  P., {Droettboom}, M., {Bray}, E., {Aldcroft}, T., {Davis}, M., {Ginsburg},
  A., {Price-Whelan}, A.~M., {Kerzendorf}, W.~E., {Conley}, A., {Crighton}, N.,
  {Barbary}, K., {Muna}, D., {Ferguson}, H., {Grollier}, F., {Parikh}, M.~M.,
  {Nair}, P.~H., {Unther}, H.~M., {Deil}, C., {Woillez}, J., {Conseil}, S.,
  {Kramer}, R., {Turner}, J.~E.~H., {Singer}, L., {Fox}, R., {Weaver}, B.~A.,
  {Zabalza}, V., {Edwards}, Z.~I., {Azalee Bostroem}, K., {Burke}, D.~J.,
  {Casey}, A.~R., {Crawford}, S.~M., {Dencheva}, N., {Ely}, J., {Jenness}, T.,
  {Labrie}, K., {Lian Lim}, P., {Pierfederici}, F., {Pontzen}, A., {Ptak}, A.,
  {Refsdal}, B., {Servillat}, M., \& {Streicher}, O. 2013, \aap, 558, A33

\bibitem[{{Bacon} et~al.(2016){Bacon}, {Piqueras}, {Conseil}, {Richard}, \&
  {Shepherd}}]{MPDAF}
{Bacon}, R., {Piqueras}, L., {Conseil}, S., {Richard}, J., \& {Shepherd}, M.
  2016, {MPDAF: MUSE Python Data Analysis Framework}, Astrophysics Source Code
  Library. \eprint{1611.003}

\bibitem[{{Bacon} et~al.(2014){Bacon}, {Vernet}, {Borisova}, {Bouch{\'e}},
  {Brinchmann}, {Carollo}, {Carton}, {Caruana}, {Cerda}, {Contini}, {Franx},
  {Girard}, {Guerou}, {Haddad}, {Hau}, {Herenz}, {Herrera}, {Husemann},
  {Husser}, {Jarno}, {Kamann}, {Krajnovic}, {Lilly}, {Mainieri}, {Martinsson},
  {Palsa}, {Patricio}, {P{\'e}contal}, {Pello}, {Piqueras}, {Richard},
  {Sandin}, {Schroetter}, {Selman}, {Shirazi}, {Smette}, {Soto}, {Streicher},
  {Urrutia}, {Weilbacher}, {Wisotzki}, \& {Zins}}]{muse2014}
{Bacon}, R., {Vernet}, J., {Borisova}, E., {Bouch{\'e}}, N., {Brinchmann}, J.,
  {Carollo}, M., {Carton}, D., {Caruana}, J., {Cerda}, S., {Contini}, T.,
  {Franx}, M., {Girard}, M., {Guerou}, A., {Haddad}, N., {Hau}, G., {Herenz},
  C., {Herrera}, J.~C., {Husemann}, B., {Husser}, T.-O., {Jarno}, A., {Kamann},
  S., {Krajnovic}, D., {Lilly}, S., {Mainieri}, V., {Martinsson}, T., {Palsa},
  R., {Patricio}, V., {P{\'e}contal}, A., {Pello}, R., {Piqueras}, L.,
  {Richard}, J., {Sandin}, C., {Schroetter}, I., {Selman}, F., {Shirazi}, M.,
  {Smette}, A., {Soto}, K., {Streicher}, O., {Urrutia}, T., {Weilbacher}, P.,
  {Wisotzki}, L., \& {Zins}, G. 2014, The Messenger, 157, 13

\bibitem[{{Bertin} \& {Arnouts}(1996)}]{sextractor}
{Bertin}, E., \& {Arnouts}, S. 1996, \aaps, 117, 393

\bibitem[{{Herenz} \& {Wistozki}(2016)}]{LSDCat}
{Herenz}, E.~C., \& {Wistozki}, L. 2016, {LSDCat: Line Source Detection and
  Cataloguing Tool}, Astrophysics Source Code Library. \eprint{1612.002}

\bibitem[{{Hutchinson} et~al.(2016){Hutchinson}, {Bolton}, {Dawson}, {Allende
  Prieto}, {Bailey}, {Bautista}, {Brownstein}, {Conroy}, {Guy}, {Myers},
  {Newman}, {Prakash}, {Carnero-Rosell}, {Seo}, {Tojeiro}, {Vivek}, \& {Ben
  Zhu}}]{hutchinson2016}
{Hutchinson}, T.~A., {Bolton}, A.~S., {Dawson}, K.~S., {Allende Prieto}, C.,
  {Bailey}, S., {Bautista}, J.~E., {Brownstein}, J.~R., {Conroy}, C., {Guy},
  J., {Myers}, A.~D., {Newman}, J.~A., {Prakash}, A., {Carnero-Rosell}, A.,
  {Seo}, H.-J., {Tojeiro}, R., {Vivek}, M., \& {Ben Zhu}, G. 2016, \aj, 152,
  205. \eprint{1607.02432}

\end{thebibliography}

\end{document}